# Finite frequentism explains quantum probability§

## Simon Saunders

Abstract: I show that frequentism, as an explanation of probability in classical statistical mechanics, can be extended in a natural way to a decoherent quantum history space, the analogue of a classical phase space. The result is a form of finite frequentism, in which Gibbs' concept of an infinite ensemble of gases is replaced by the quantum state expressed as a superposition of a finite number of decohering microstates. It is a form of finite and actual (as opposed to hypothetical) frequentism insofar as all the microstates exist, even though they may differ macroscopically, in keeping with the decoherence-based Everett interpretation of quantum mechanics.

## 1. Introduction

Frequentism in philosophy of probability holds that the probability of an event is its relative frequency of occurrence in a suitable ensemble. Of its many champions, perhaps the most prominent was von Mises. For him the ensembles were finite and actual, but as such they were vulnerable to 'the reference class problem' -- the problem of *which* ensemble, as generally there are either too many, or too few. Von Mises himself recommended pluralism: there is no such thing as 'the' probability of an event, only probabilities relative to different references classes, in general contradicting each other – and for too rare an event, it may be, no probability at all (as lacking a reference class). But that proved unpersuasive. Finite frequentism as a philosophy of probability is today routinely dismissed as unworkable.[1]

However, the relevant ensembles in von Mises' sense were collections of events in a classical world. In the real word we find the quantum. Quantum mechanics is the most prolific, most accurate, and most unificatory theory in all the long history of physics, but also, astonishingly, now after more than a century, the most controversial. It is seemingly incapable of a realist reading -- hence the repeated attempts to add to or modify the formalism, despite its empirical success. The main difficulty is 'the quantum measurement problem', which concerns the way probability is usually introduced into quantum mechanics, using the Born rule: a rule that puzzlingly seems always to involve the notion of 'observer' and 'observation', however indirectly.

The one clear realist reading of standard quantum mechanics takes seriously the idea that the superposition principle has unlimited validity, and that the Schrödinger equation can thus be applied to macroscopic bodies. It also takes seriously the reality of quantum states, including when entering into superpositions. It is otherwise conservative, leaving quantum mechanics unchanged. The result is realism about superpositions of states that differ macroscopically, as arise in particular following quantum measurements. With no theoretical resource for picking out one rather than another outcome as actual, all the outcomes are actual, entering a grand superposition -- this the reasoning that leads to many worlds, otherwise known as the Everett interpretation of quantum mechanics, first introduced in (Everett [1957]).

---

§ To appear in the *British Journal for the Philosophy of Science*, forthcoming.

[1] As in (Rowbottom [2015 p.112]), (La Caze [2016]), (Myrvold [2016, 2021]). A particularly influential critique is (Hájek [1996]). For a dissenting voice, see (Hubert [202]). (Here I put to one side Humean 'best-system' accounts of probability, which are sometimes considered sophisticated forms of actual frequentism.)

No wonder, if this is on anything like the right lines, the controversy over quantum mechanics has been unending. Once formed, short of super-technology or Poincaré recurrence, the states that arise on measurement, and the states that arise from them in turn on further measurements, will never interfere with one other. But this is the only way states *can* interact with one another, in a linear theory; so once formed, each state evolves as if all the others were not there. This explains the appearance of quantum jumps and collapsing wave-packets.

The picture is challenging to say the least, but the question is whether it succeeds on its own terms, and nowhere more so than when it comes to probability. The Born rule involves the squared amplitudes of macrostates produced by measurements: It is not obvious how, with all else unchanged, differences in amplitudes yield differences in probabilities. Attention in recent years to the decision-theory approach to probability (Deutsch [1999], Wallace [2010, 2012]) has for some only reinforced the worry: for if the Born rule is explained in terms of rational agency – if it is, indeed, rationally forced, as claimed, for agents knowledgeable of the superpositions of macrostates produced by measurements -- then it may be that entirely accounts for probability, with no residue to be explained.[2] Or rather, there remains one small residue: the relative frequencies of macrostates (experimental outcomes) on repeated measurements. But a rule based on these, as in (Graham [1973]), contradicts the Born rule, and so is of no use in explaining quantum probability. On the contrary, it creates only a further difficulty.

The situation is transformed when branches are defined in decoherence theory.[3] Branching, still in Everett's sense, for macroscopic bodies, now takes place at the level of differences in molecular configurations and momenta. At ordinary temperatures and pressures, in ordinary matter, branching takes place all the time, and with extreme rapidity (much faster than thermal relaxation times). It is taking place in an experiment even if the apparatus just sits there and doesn't measure anything -- reflected only in all those minute differences in thermal molecular motions, invisible to the eye.

From this perspective, the old branch-counting rule is simply an aberration: the point is to count the relative numbers of microstates for the various experimental outcomes, not the numbers of outcomes. And yet, until very recently, no new rule was forthcoming based on the corrected idea. To the contrary, standard wisdom has it that decoherence is an inherently messy, approximate affair, and that branches themselves are emergent entities, too poorly defined to have any precise boundaries – so that ultimately, branch number makes no physical sense.[4]

---

[2] According to (Deutsch [2016]), (Brown and Porath [2020]), this is a virtue of the approach, not a deficiency.

[3] Following (Joos and Zeh [1985]), (Tegmark [1993]), (Saunders [1993]); it was hinted in (Gell-Mann and Hartle [1990]).

[4] As argued, among others, by me, in (Saunders [2005]). See also (Wallace [2012 pp.99-100]).

Yet this consensus may be challenged. The theory of decoherent histories, as introduced in (Gell-Mann and Hartle [1990]), provides a common language for decoherence theory, including for master equations for the reduced density matrix.[5] It further *does* permit a precisification of microstates, cunningly shifting the approximation away from the states and onto the no-interference condition: branches may be well-defined, yet only approximately decohere (and residual but tiny interference is just fine for Everettians, as may usefully lead to observable effects). The upshot: there may be no point at which a unique notion of branch number can be defined, but well away from that point, *ratios* in branch numbers may yet make physical sense.

It is more than a hope. As shown in (Saunders [2021]), when the decoherent history space involves a continuous dynamical variable (like position or momentum), a branch-counting rule can be defined that for finite ensembles approximately agrees with the Born-rule. The rule counts microstates parameterised by histories of events up to time $t$: the total number, up to that time, and those that contain event $E$, at some earlier time $t' \leq t$; the ratio of the two is then the probability (relative frequency) of $E$. The rule has a ready extension to an infinite ensemble by taking the limit $t \to \infty$ (assuming branching is unending): in that limit the relative frequency of $E$ equals the Born rule probability of $E$, to which relative frequencies for finite $t - t'$ are approximations (with what I called 'rounding errors'). The rule is, moreover, traceable to Boltzmann's microstate counting rule on classical phase space, so it is scarcely ad hoc, replacing Liouville measure by the Hilbert-space norm.

There are, however, certain negatives to this approach. Its dependence on the infinite-time limit and hence cosmology is troubling: is there no precise notion of quantum probability in a re-collapsing universe, subject to Poincaré recurrence? If so, it is not obvious, absent the infinite time limit, what 'approximation' (and 'rounding error') really mean -- or not in terms of frequentism. One might further have thought the physical chances for local events are local to those events. And the parallel between Liouville measure and the Hilbert-space norm is hardly new; it was first drawn in (Everett [1957]).

In what follows I propose a different strategy: a purely synchronic microstate-counting rule, restricted to finite ensembles throughout. Decoherence is still important (and remains an intrinsically diachronic concept), but for our purposes its role can be limited to the definition of the parameter space (fixing the decoherence basis). The comparison with the methods used in classical statistical mechanics is now direct – and shows that the relevant parallel is not between the Hilbert space norm and Liouville measure, but between the Hilbert space norm of the wave-function in the decoherence basis and Gibbs' concept of 'density-in-phase', independently specified, as defining the ensemble.

Boltzmann's rule, accurately taken over to quantum mechanics, yields a new and different branch-counting rule that we will also consider. It is similar to Graham's rule in treating 'non-zero amplitude' as a sufficient as well as a necessary criterion for the existence of branches: a branch is a certain sort of state, and it exists if and only if it has non-zero amplitude (whereas on the Gibbs approach, as we shall see, the amplitude is used to define the branch). But however intuitive, the rule fails as a synchronic probability rule. Simple alternatives are also considered and similarly fail. They fall victim to what remains of the reference-class problem – the problem of reconciling different relative frequencies for an experimental outcome when referred to different but equally permissible ensembles of microstates, as produced on measurements.

---

[5] See (Kiefer [1996]) for an overview. For realistic cases, involving macroscopic numbers of particles, decoherent history spaces are 'quasiclassical domains', in Gell-Mann and Hartle's sense, depending on the Hamiltonian, degrees of freedom involved, and initial quantum state.

In contrast, the Gibbs rule passes this test. The relative frequencies defined by distinct admissible ensembles necessarily agree. They agree too with the Born rule. The ensembles are invariably finite. Finite frequentism would thus appear to explain quantum probability quite generally – and, in view of the universality of quantum mechanics, to explain probability in the objective or physical sense in its entirety.

Or so I claim, embracing realism about decohering microstates, when the ensembles involved are actual as well as finite. Different ensembles are different ways of parcelling up the same total state, with respect to the same decoherence basis, into finitely-many parts. Might antirealists about macroscopic superpositions make use of this method as well? For them, like Gibbs, the ensembles will be imaginary, but now rendered finite: that may be viewed as an improvement on hypothetical frequentism, where the ensembles are usually infinite. But as we shall see, permissible ensembles, for any event $E$, will in general include Schrödinger-cat states for $E$ – states that are superpositions of microstates in which $E$ occurs, and microstates in which $E$ does not occur. Because included in ensembles, such states have non-zero relative frequency, and so non-zero probability. That is unlikely to be acceptable to those who reject macroscopic superpositions, since presumably such states then have a chance of being actual. But there is a possible extension of the Gibbs quantum rule to contexts where decoherence does not occur (at the microscopic level proper), and that may prove of interest to sceptics. It is discussed in §8.

Pilot-wave theory (otherwise known as Bohmian mechanics) adds additional 'hidden' degrees of freedom, but otherwise preserves the Schrödinger equation with the same initial state: it therefore contains the same ensembles of decohering microstates. It does both better and worse than antirealism: better, in that those microstates have some kind of reality, if only as guiding ('piloting') the hidden-variables; worse, in that on introducing those additional variables, once assigned definite values, the statistical patterns we are considering become only a part of a larger physical picture, that may tell a quite different story about probability.[6]

A last question, now, for the Everett interpretation. Might the new rule, in avoiding the usual pitfalls of finite frequentism, be too good to be true? Perhaps it isn't a form of frequentism at all, but a new and sui generis theory of physical probability. The epistemology, in particular, seems quite different from anything in von Mises' writings, as the ensembles are in practise unobservable.

I refuse to be so flattered. The origins of the new rule lie squarely in the ideas of Boltzmann and Gibbs, and they were surely frequentists. They employed imaginary, infinite ensembles, so for them too the ensembles were unobservable -- indeed, unobservable in principle. Others too, of that period, considered ensembles of possibilities, again unobservable in principle.[7]

---

[6] For example, a non-equilibrium story, as introduced by Valentini in his ([1991]), his reason too for rejecting the 'Everett in denial' criticism of pilot-wave theory (Valentini [2010]). On the general point, see (Hawthorne [2010]).

There follow two pieces of stage-setting, first on frequentism in classical statistical mechanics, and the second on decoherent histories theory in quantum mechanics. Readers familiar with both may jump directly to §4, where the two quantum rules are defined. §5 introduces the key idea of 'interval probability', replacing the notion of approximate probability, and defined in purely finitary terms. With that it is possible to define general consistency requirement. The Boltzmann rule is shown to be inconsistent in this sense, whereas the Gibbs rule passes this test, shown in §6 and §7 respectively. The two remaining sections are more speculative, and given over to more general philosophical questions, including the epistemology of probability.

## 2. Microstate counting in classical statistical mechanics

The key idea in (Boltzmann [1877]) was to partition the accessible phase space for a gas into a finite number of cells, all the same size, and to consider the relative numbers of such cells consistent with one macrostate (in which the gas has a certain pressure and volume, for given total energy), in comparison with another. The macrostate with the greatest number relative to all the others is the equilibrium state – therefore, according to frequentism, the most probable state. In this way Boltzmann was able to compute the equilibrium entropy and obtain the equation of state for the ideal gas.

In this the concept of equiprobability and ensemble probability fitted together seamlessly; if the probability of $E$ is the relative frequency of $E$ in an ensemble, the ensemble elements must all have the same probability; and conversely, if all the elements in an ensemble have the same probability, then the relative frequency of E in the ensemble must be its probability.

The size of a cell $\gamma_k \subseteq \Gamma$, where $\Gamma$ is classical phase space, is given by its Liouville measure $\sigma: \Gamma \to \mathbb{R}^+$. For $N$ particles in rectilinear coordinates it is:

$$\sigma(\gamma_k) = \int_{\gamma_k} dp_1 \dots dp_{3N} dx_1 \dots dx_{3N}. \tag{1}$$

To ensure that the choice of partitioning $\{\gamma_k\}$ be irrelevant, Boltzmann took the limit $\sigma(\gamma_k) \to 0$ in which microstates are points in $\Gamma$. This solved the reference-class problem, at the price of passing over to infinite (and of course imaginary) frequentism.

This idea of probability as relative frequency proved immensely important to the subsequent history of discovery of quantum mechanics, first in the work of Planck, in his attempt to ground the new black-body radiation law on the Boltzmann entropy, as defined by ratios in numbers of microstates, and then in Einstein's 1905 calculations, that followed Planck in deriving the equilibrium entropy density from the radiation law, and went on to show that it behaved under fluctuations in the same way as the entropy of an ideal gas (but working only from the Wien limit of Planck's law). Einstein later, in one of the few places

---

[7] For example, von Kries ([1886]). Everett's branches have also been understood in terms of possible worlds (Wilson [2020], although embracing a form of modal realism), and in terms of perspectives on a single world (Barrett [2011]).

where he returned to this early history of quantum theory, highlighted the importance of Boltzmann's ideas:

> On the basis of the kinetic theory of gases Boltzmann had discovered that, aside from a constant factor, entropy is equivalent to the logarithm of the "probability" of the [macro]state under consideration. Through this insight he recognized the nature of the course of events which, in the sense of thermodynamics, are 'irreversible'. Seen from the molecular-mechanical point of view, however, all courses of events are reversible. If one calls a molecular-theoretically defined state a microscopically described one, or, more briefly, microstate, then an immensely large number of states belong to a macroscopic condition. [This number] is then a measure of the probability of a chosen macrostate. This idea appears to be of outstanding importance also because of the fact that its usefulness is not limited to microscopic description on the basis of [classical] mechanics. (Einstein [1949 p.43]).

What is more, the idea led to the discovery of quantum mechanics precisely by *not* taking the $\sigma(\gamma_k) \to 0$ limit (this was Planck's daring step in 1900, departing from Boltzmann's method). It was finite frequentism indeed, albeit imaginary.[8]

Boltzmann's ideas are today still familiar, but mainly for what they say about entropy. Except for special cases, like equilibrium states for finite-dimensional Hilbert spaces, the link with frequentism has been ignored. Classically, microstate-counting is seen as little more than a calculational device; conceptually, since dependent on a volume measure $\sigma$ on $\Gamma$, it seems more straightforward to identify the thermodynamic probability of a macrostate outright with ratios in volumes. But the very term 'thermodynamic probability' has become suspect; as argued in (Albert [2000]), Liouville measure enters in two quite distinct ways, the first in defining the Boltzmann entropy of a given macrostate (as the logarithm of the accessible phase-space volume), and the second in defining a probability distribution over microstates. The sense in which the Boltzmann entropy of a closed system will only *probably* increase is made out with respect to the second. And even talk of 'probability' is in question, to be replaced by the concept 'typicality', especially in the adjacent literature on probability in pilot-wave theory. Everett too spoke of typical branches and observers rather than probable ones.[9] In contrast, moving to the quantum analogue of Boltzmann's rule, we hold fast to the notion of a finite partitioning of state space. The great improvement, conceptually, is that the ensembles thus defined are actual, rather than imaginary.

Now for the background to our second rule, based on Gibbs' concept of probability. Like the Gibbs entropy, this is defined in terms of a new notion introduced on phase space, foreign to Boltzmann's writings – the *density-in-phase* function $D: \Gamma \to \mathbb{R}^+$. This was used to define a probability density on $\Gamma$ (what Gibbs called 'the coefficient of probability'). The approach was still frequentist -- it was Gibbs who coined the term 'statistical mechanics' – but abjured appeal to finite ensembles (we will modify his approach accordingly, to bring it into line with Boltzmann's procedure). He defined various equilibrium ensembles, each with its density-in-phase function (microcanonical, canonical, and grand canonical), depending on the external constraints (on total energy, temperature, and particle number respectively). He was explicit on the status of members of these ensembles: they were 'creatures of the imagination' (Gibbs 1902 p.188); so again this was hypothetical (imaginary, infinite) frequentism.

---

[8] This involved the dimensionality of certain finite-dimensional Hilbert spaces, and a completely degenerate initial state (density matrix). Photon indistinguishability greatly complicated this history: see Saunders [2020a], [2020b].

[9] Emphasised in (Barrett [2016]); for a recent defence of Everett's treatment of probability, see (Lazarovici [2023]). Typicality was introduced in pilot-wave theory in (Dürr et al [1992]); see also (Goldstein [2001]), (Myrvold [2016]).

If we follow Boltzmann's procedure by coarse-graining a Gibbs ensemble into a finite collective of equiprobable microstates there is now a clear alternative to the equivolume rule. Microstates should be equiprobable using the density-in-phase function $D$. We go over to finite ensembles of equiprobable microstates by choosing a partitioning $\{\gamma_k\}$ of phase space in such a way that each cell has the same fraction of the ensemble, as determined by $D$ – that is, equality in the quantities:

$$\int_{\gamma_k} D(q) d\sigma \qquad (2)$$

rather than equality in (1) – or more precisely, since Gibbs thought the numbers involved may all be infinite, equality in the ratios of (2) with the total number $N$ of systems in the ensemble.

Here, in his own words, is how Gibbs made the connection with probability:

> Now, if the value of $D$ is infinite, we cannot speak of any definite number of systems, within any finite limits, since all such numbers are infinite. But the ratios of these infinite numbers may be perfectly definite. If we write $N$ for the total number of systems, and set
>
> $$P = \frac{D}{N},$$
>
> $P$ may remain finite, when $N$ and $D$ become infinite. The integral
>
> $$\int P d\sigma$$
>
> taken within any given limits, will evidently express the ratio of the number of systems falling within those limits to the whole number of systems. This is the same thing as the *probability* that an unspecified system of the ensemble (i.e. one of which we only know that it belongs to the ensemble) will lie within the given limits. The product $P d\sigma$ expresses the probability that an unspecified system of the ensemble will be found in the element of extension-in-phase $d\sigma$. We shall call $P$ the *coefficient of probability* of the phase considered. (Gibbs [1902 p.16]].

The procedures of Boltzmann and Gibbs give the same results for the equilibrium states of isolated gases of non-interacting particles, like ideal gases and black-body radiation (when $D$ is uniform), but they differ for non-ideal material gases and for non-equilibrium systems, of the sort relevant to quantum measurements.

## 3. Decoherent history space

A decoherent history space is a structure $\langle \mathcal{M}, \lambda, |\psi\rangle, \widehat{H}, \mathcal{H} \rangle$, where $\mathcal{M}$ is a parameter space, the analogue of phase space (in the simplest case defined in terms of the spectra of some set of commuting self-adjoint operators), $\lambda$ is an additive, dimensional measure on $\mathcal{M}$ (the analogue of Liouville measure), $\widehat{H}$ is the Hamiltonian, containing the dynamics, and $|\psi\rangle \in$ $\mathcal{H}$ is the total ('universal') state in the Hilbert space $\mathcal{H}$.[10] We further assume that $\mathcal{M}$ includes the spectrum of at least one continuous operator (like position or momentum). This is physically reasonable; all events take place in space, and the spectrum of the position operator is continuous. We further assume $\mathcal{M}$ is bounded, $\lambda(\mathcal{M}) < \infty$.

More is needed to define quantum histories, notably a discretisation of the time, $\{t_k\}, k = 1, \ldots, N$, from which $N$-step histories can be defined. But we will not need histories here: our approach is synchronic, and we make do with the structure $\langle \mathcal{M}, \lambda, |\psi\rangle \rangle$. This is to be compared to a classical probability space $\langle \Gamma, \sigma, \mu \rangle$, where $\Gamma$ is an event space, equipped with an additive measure $\sigma$ (this could be phase space and Liouville measure respectively, so we use the same symbols) and an additive probability function $\mu$.

As before we call the elements $\{\alpha_k\}$ of a partitioning 'cells', meaning they are disjoint ($\lambda(\alpha_j \cap \alpha_k) = 0$ for $j \neq k$), and they cover or tile $\mathcal{M}$ ($\bigcup_k \alpha_k = \mathcal{M}$). For the continuous part of $\mathcal{M}$, we assume cells are connected. In comparison to the classical case, the main novelty is the canonical mapping from $\alpha \subseteq \mathcal{M}$ to states $P_\alpha |\psi\rangle \in \mathcal{H}$, where projectors $P_\alpha$ and the measure $\lambda$ come by construction of $\mathcal{M}$ (via Stone's theorem) from commuting operators. In this way algebraic operations involving projectors mirror set-theoretic operations on $\mathcal{M}$: for disjoint $\alpha, \alpha'$, the projector onto their union equals the sum of their projectors:

$$\alpha \cap \alpha' = \emptyset \Rightarrow P_{\alpha \cup \alpha'} = P_\alpha + P_{\alpha'} \qquad (3)$$

and projectors onto intersections of $\alpha, \alpha'$, now without restriction, go over to products:

$$P_{\alpha \cap \alpha'} = P_\alpha P_{\alpha'} = P_{\alpha'} P_\alpha \, .$$

It follows that if $\alpha, \alpha'$ are disjoint, the states $P_\alpha|\psi\rangle$, $P_{\alpha'}|\psi\rangle$ are (exactly) orthogonal. For a classical probability space $\langle \mathcal{M}, \sigma, \mu \rangle$, in place of the canonical map $\alpha \to P_\alpha$ we have the map $\gamma \to \mu(\gamma)$ of cells in $\Gamma$ to probabilities. Eq.(3) then goes over to additivity of $\mu$, and has to be assumed separately.

Consider now a finite partitioning $\alpha_1, \alpha_2, \ldots, \alpha_s$ of $\mathcal{M}$, (were $\mathcal{M}$ a discrete space, these could be eigenvalues of the associated operators). Since $P_\mathcal{M}$ is the identity on $\mathcal{H}$, it follows, at any time $t$:

$$|\psi\rangle = P_{\bigcup_k \alpha_k}|\psi\rangle = \sum_{k=1}^{s} P_{\alpha_k}|\psi\rangle \, .$$

In this way the total state $|\psi\rangle$ at $t$ is represented as a superposition of finitely-many states $P_{\alpha_k}|\psi\rangle$, complete with their amplitudes and phases. The state $|\psi\rangle$ we are interested in is that shortly following a measurement process, evolved under the Schrödinger equation, where states $P_{\alpha_k}|\psi\rangle$ may differ macroscopically (in particular, as to experimental outcome). However, they may not be decohering states: for this some coarse-graining is always necessary. We assume this of our partitioning $\{\alpha_k\}$ until further notice. These are the ingredients, pending further specification of the partitioning, of our two microstate-counting rules.

---

[10] We use the notation '$|\psi\rangle$', for (vector) states, but make no assumption of normalisation. Viewed realistically, in the Everett interpretation as in pilot-wave theory, the amplitudes of branch states are always rapidly decreasing in time (only relative amplitudes have a physical meaning, as likewise phase).

The one further input still needed, from the underlying physics, is that in macroscopic experiments decohering microstates may provide considerable microscopic detail at molecular, and even at atomic scales, for ordinary matter at ordinary temperature. The development of superpositions of orthogonal states like these over time under the Schrödinger equation will be rapid and quite uncontrollable.[11] Quantum jumps, in other words, are everywhere. There are of course limits, as Schrödinger, in this Journal, famously argued[12], but that is only to say some coarse-graining of $\mathcal{M}$ is needed. At the macroscopic level, then, the available ensembles will be large, of the order of Avogadro's number. But we need to make sense of small ensembles too, for coarse-graining always preserves decoherence; relative frequencies as defined in small ensembles (where the cells may be macrostates) had better make sense, just as in large ones, so long as the ensembles conform to the same definite rule. But which rule?

## 4. Quantum microstate-counting rules

In light of §2, there are the two obvious candidates rules for defining ensembles of microstates: one, following Boltzmann, based on the idea of cells of equal Liouville measure at an instant in time, accessible to the individual gas; and the other, following Gibbs, based on the idea of an ensemble endowed with a density-in-phase function $D$, at an instant of time, and cells containing equal fractions of the ensemble, as defined by $D$.

It might be thought that there is a third option: why not count branches simpliciter, without appeal to a partitioning of $\mathcal{M}$ at all? But the idea is confused. At best it amounts to 'count non-zero eigenstates of some self-adjoint operator'. But as noted, even were $\mathcal{M}$ a discrete manifold, with the $P_{\alpha_k}|\psi\rangle$'s eigenstates of operators, they will not in general decohere. Some coarse-graining is needed. As Wallace puts it:

> [T]here is no sense, in which these phenomena lead to a naturally discrete branching process: as we have seen in studying quantum chaos, while a branching structure can be discerned in such systems, it has no natural "grain". To be sure, by choosing a certain discretisation of (configuration-) space and time, a discrete branching structure will emerge, but a finer or coarser choice would also give branching. And there is no "finest" choice of branching structure: as we fine-grain our decoherent history space, we will eventually reach a point where interference between branches ceases to be negligible, but there is no precise point where this occurs. (Wallace [2012 p.99-100]).

Counting numbers of eigenstates, with $\mathcal{M}$ a purely discrete manifold, is not to count decohering states. In any case, in realistic cases, $\mathcal{M}$ involves a continuous space, so there is no 'natural grain'; that is our assumption here.

---

[11] For quantitative estimates, see (Tegmark [1993]) and, for a case of special interest (neuron-firing), (Tegmark [2000]). The states that decohere are spatially localised states of molecular centre-of-mass degrees of freedom (hence the detail at the molecular level). This is because entanglement with the environment at the relevant ('fast') timescale is produced by local (electromagnetic) couplings (Zurek [1981]).

[12] 'Quantum jumps' cannot be occurring at the level of coherent interactions between molecules, involving individual electrons and other light particles, for they are needed to explain the physics of ordinary matter (Schrödinger [1952]):

We have the two methods as stated. They each promise a reductive account of probability insofar as the reductive base is defined independent of probabilistic concepts. For the first method, where the cells are fixed in size, we suppose the idea 'accessible cell' is made out in the terms 'cell of non-zero amplitude'. This seems the most natural requirement, once the cells are fixed, and it is the one used in Graham's branch-counting rule, when the cells were macrostates (measurement outcomes). Since the cells all have the same finite size, and given that $\mathcal{M}$ is bounded, it follows that such ensembles are always finite.

For the second method we need the analogue of Gibbs' ensemble and its associated density-in-phase function. To see what this should be, observe that we are replacing Gibbs' infinite collection of classical (point-like) microstates, each the state of an imaginary gas, equipped with the function $D$ on $\Gamma$, by an ensemble of microstates complete with their amplitudes and relative phases, comprising -- superposing to give -- the total quantum state, that we may write as a wave-function on $\mathcal{M}$. Rather than dividing up the classical ensemble into microstates with equal fraction of the ensemble, as defined by the density in phase, *we divide the quantum state into microstates each with equal fraction of the total state*, as defined by the wave-function. That is, we divide up the quantum state into branches of equal amplitude.

Equivalently, if an additive quantity is wanted (corresponding to a density which can be integrated, as in (2)), so that the fractions add, we divide the quantum state into microstates of equal *squared* amplitude. But the two methods are the same, as amplitudes are non-negative real numbers.

Gibbs' density in phase as a non-negative function on $\Gamma$ is thus identified with the squared norm of the quantum state as a function on $\mathcal{M}$. Whereas for an infinitesimal neighbourhood $d\sigma$ of a point $q$ in phase space, $D(q)d\sigma$, divided by $N$, is the fraction, or relative number, of the ensemble in the neighbourhood $d\sigma$ of $q$, we now have $|\psi(q)|^2 d\lambda \coloneqq |\langle q|\psi\rangle|^2 d\lambda$, where $q \in \mathcal{M}$, divided by the squared amplitude of $|\psi\rangle$, as the fraction, or relative amount, of the wave-function in the neighbourhood $d\lambda$ of $q$.

Branches thus defined had better correspond to cells of non-zero measure $\lambda$. This follows more or less automatically: if $\lambda(\alpha_k) = 0$ then $P_{\alpha_k}|\psi\rangle = 0$ for any reasonable volume measure on $\mathcal{M}$. But it gives pleasing symmetry to the two branch-counting rules to make this explicit: the one, following Boltzmann, counts cells of equal volume that have non-zero amplitude, whilst the other, following Gibbs, counts cells of equal amplitude that have non-zero volume.

With that it is clearer that *both* rules are in question as reductive accounts of probability in quantum mechanics. For of course 'amplitude' in ordinary quantum mechanics *does* have a connection with probability, through the Born rule -- the only place where probability enters the picture. 'Non-zero amplitude' means 'non-zero probability', and 'equiamplitude' means 'equi-probable'.

Yet that is not quite right, as on the Boltzmann rule, equivolume cells with non-zero amplitude are equiprobable, not just of non-zero probability. But the Gibbs rule looks more vulnerable. For a macrostate $\beta$, the quantity

$$\frac{\int_\beta^{\square} |\psi(q)|^2 d\lambda}{\int_\mathcal{M}^{\square} |\psi(q)|^2 d\lambda} \qquad (4)$$

is just the Born-rule probability for $\beta$; it corresponds to Gibbs' expression:

$$\frac{\int_\beta^{\square} D(q) d\sigma}{\int_\Gamma^{\square} D(q) d\sigma} \; . \qquad (5)$$

As Gibbs said, this is the probability that an unspecified system of the ensemble will be found in $\beta$. Yet recall *why* he said that: it is because (5) 'will evidently express the ratio of the number of systems falling within those limits to the whole number of systems', and immediately following, '….this is the same thing as the *probability*….'. The interpretation of the expression (5) as a probability is to *follow* from frequentism, it is not assumed by it. If, indeed, it made sense to speak of the wave-function as made up of infinitely many parts, the argument in the quantum case would be identical. Instead, we break it up into finitely-many equal parts, entities given, not sui generis, but by construction. We are building up to a notion of probability, not building one in; only later, and then in a qualified sense, will be able to recover an interpretation of Eq.(4) in terms of probability.

As for the concept 'amplitude', it is clearly intelligible in contexts that have nothing to do with probability. Like phase, it is a new physical primitive of quantum theory. But there remains the branch concept itself ('decohering microstate'). Does decoherence theory depend, in its formulation or derivation from the Schrödinger equation, on probabilistic reasoning or justification? It is certainly usually *interpreted* in terms of probability, and its practitioners, when pressed as to the meaning of given expressions, may well appeal to the Born rule and operational definitions when pressed. The move may be safe, but it is not always explanatory, not least of one of the central notions of decoherence theory: the concept 'interference' itself. This was never happily explained in terms of 'probability waves', whatever *they* might be, but it is immediately intelligible in the cancelling or reinforcing of categorical physical quantities that take positive and negative values, as in any wave theory involving amplitude and phase. We do not need probability to understand interference in quantum theory. We should distinguish interference from an interference *effect*, whereby interference is made observable, so that measurement, arguably, is involved; that, it may be granted, in ordinary quantum mechanics, is to make explicit or implicit use of the Born rule; but (at least for realists about states) interference takes place at the locus of the screen even when the screen is removed and no observation is made.

To say more on this topic would take us in two different directions, one technical, involving more decoherence theory, and one philosophical, and the need in the physical sciences, if it is a need, for operational justifications for approximations, model-building, and representation. I forego both expeditions, noting that the same issues arise for the decision-theoretic argument in (Wallace [2010, 2012]), which first attracted this kind of criticism.[13]

I conclude there is a prima facie case that 'amplitude' and 'branch state' are intelligible physical notions that need have nothing to do with probability. It follows that *both* branch-counting rules promise to provide reductive theories of chance. However, they cannot both be true (except, perhaps, in special cases), and perhaps neither is satisfactory.

---

[13] For example, by Kent [2010], Dawid and Thèbault [2015]; for replies see Saunders [2022], Franklin [2024].

## 5. Consistency

Because of the decoherence condition, on either the Boltzmann or Gibbs rule we are limited not just to finite partitionings of $\mathcal{M}$ at any time (that follows from $\lambda(\mathcal{M}) < \infty$ and $|||\psi\rangle|| < \infty$ respectively), but to partitionings that cannot be arbitrarily large. Very often in the sequel we will work with small ensembles. But there is no conflict with Kolmogorov's axioms. On the contrary, probabilities defined as relative frequencies in a finite ensemble are automatically additive, normalised, and contained in the interval [0,1]. As one of the many attractions of frequentism, Kolmogorov's axioms are derived and not assumed.[14]

However, having defined the kind of ensemble (equiamplitude or equivolume), there are still infinitely many ensembles available, as we are assuming $\mathcal{M}$ varies continuously. It is what remains of the reference-class problem: to which ensemble, exactly, should a macrostate be referred? Our strategy is similar to von Mises': *any* admissible ensemble should be acceptable, as can be represented in the state following a measurement, and should yield, for any macrostate, a perfectly meaningful relative frequency -- save only that some may be more informative than others (we shall see how this works in a moment). The crucial difference from von Mises' approach is that we further require that all relative frequencies be mutually consistent with one another.

To make this precise, we need to use a definite rule, as the details vary in the two cases -- but not by much, so it will be enough to use just one rule; the arguments are similar, and the consistency conditions we end up with are the same. We use the equivolume rule.

Suppose $\{\alpha_k\}$ is a partitioning of $\mathcal{M}$ into $s$ cells, $n$ of which have non-zero amplitude, all with $\lambda(\alpha_k) = \tau$. Let macrostate $\beta$ represent an experimental outcome (say the configuration of a Stern-Gerlach apparatus, on obtaining the outcome 'spin-up'), and suppose further that $\beta$ is exactly partitioned by $\{\alpha_k\}$ -- that is, for some index set $K$, we have $\beta = \bigcup_{k \in K} \alpha_k$, where $K$ has cardinality no greater than $s$. Then for any state $|\psi\rangle$ there will be an integral number $n_\beta$ of microstates in $\{P_{\alpha_k}|\psi\rangle; k \in K\}$ of non-zero amplitude, out of $n$ in total, $n_\beta \leq n$. The probability of $\beta$ is then (exactly) $n_\beta/n$.

Our first consistency condition is that (for the same state $|\psi\rangle$ and decoherent history space) *every* admissible partitioning that exactly partitions $\beta$ yields the same ratio. If on a new exact partitioning of $\beta$ there are $n_\beta'$ spin-up branches out of $n'$ in total, we require $n_\beta/n = n_\beta'/n'$.

The question now arises: What if $\beta$ is *not* exactly partitioned by the $\{\alpha_k\}$? Indeed, unless $\lambda(\beta)/\tau$ is an integer, there can be no index set $K$ with the required properties. One might hope to vary $\tau$ to obtain an exact partitioning of $\beta$, but that trick only works once; since $\mathcal{M}$ is exactly partitioned, this is only possible if $\lambda(\beta)/\lambda(\mathcal{M})$ is rational. See Fig.1.

---

[14] Admittedly *countable* additivity, Kolmogorov's main innovation, requires more, but we make no use of that here.

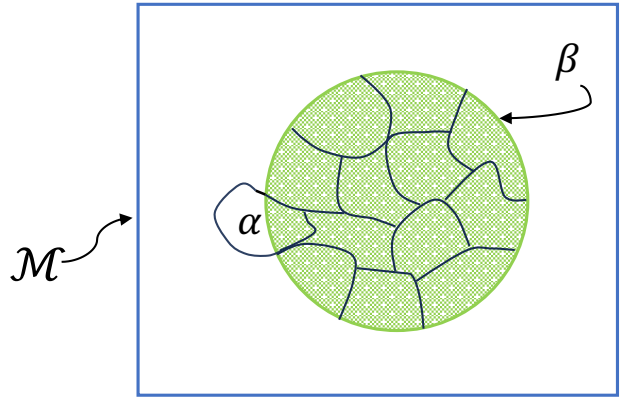

Figure 1

A bounded 2-dimensional continuous state space $\mathcal{M}$ and macrostate $\beta$, stippled. Only cells intersecting with $\beta$ are shown. On any equi-area partitioning of $\mathcal{M}$, whatever the shapes of the cells, at least one (as shown, the cell $\alpha$) will straddle the boundary of $\beta$, unless the ratio of the areas of $\mathcal{M}$ and $\beta$ is rational.

Failing this condition, evidently $\beta$ cannot be exactly partitioned whatever the shapes of the cells – some cells will neither be contained $\beta$, nor in its complement. If of non-zero amplitude, the associated microstates will be superpositions of $\beta$ and its complement: we shall say they are 'indefinite on $\beta$'.

We assume macrostates in $\mathcal{M}$ are in general defined independent of any partitioning. Is the accompanying inevitability of indefinite microstates a difficulty? But indefiniteness per se is hardly novel, for it arises in a classical probability space $\langle \Gamma, \sigma, \mu \rangle$. If we understand events in terms of sentences, a union of disjoint cells goes over to a disjunction of contradictory sentences, and intersections go over to conjunctions. Any sufficiently coarse-grained description will fail to specify some macroscopic property (will be indefinite on that property) – meaning, it will be disjunctive in the relevant sense. But further, classically, the disjuncts will be assigned exact probabilities under $\mu$, and the sum of the probability of the disjuncts will equal the probability of the disjunction, as follows from the additivity of $\mu$ at that finer-grained level.

In the quantum case there is no fixed, background probability measure $\mu$, and we have only families of finite ensembles, families of relative-frequency probability distributions. But we require the same condition: the probability assigned to a cell $\alpha$ in one ensemble, indefinite on $\beta$, where $\alpha = \gamma \cup \gamma'$ and $\gamma, \gamma'$ are both definite on $\beta$, should be the sum of the probabilities of $\gamma, \gamma'$, as defined by any other admissible ensembles. In terms of the associated microstates:

$$P_\alpha |\psi\rangle = P_\gamma |\psi\rangle + P_{\gamma'} |\psi\rangle$$

the probabiliy of the microstate on the LHS (a Schrödinger-cat microstate for $\beta$) should be the sum of the probabilities of the states on the RHS (a live-cat state, a dead-cat state), supposing they have exact probabilities, each in some ensemble.

Observe further that were there a unique, finest partitioning of $\mathcal{M}$, from which all others are constructed by sums, the consistency condition would follow automatically from additivity of probabilities at that level -- with the latter, in turn, following from frequentism. The consistency condition requires more only if there is no such finest, unique, partitioning. When $\mathcal{M}$ contains continuous parameters there is none such[15] – and note, for future reference, that this would still be the case even if we removed the constraint that microstates decohere, and allow cells $\alpha_k$ to have arbitrarily small sizes, so long as $P_{\alpha_k}|\psi\rangle \in \mathcal{H}$.

The key point is that there is nothing 'incorrect' or 'approximate' in the probability assigned to a cell $\alpha$ indefinite with respect to some macrostate $\beta$. If $P_\alpha|\psi\rangle$ is itself a microstate in an admissible ensemble of $n$ microstates, its probability is $1/n$, end of story, whether or not it is a Schrödinger-cat state for some macrostate. The difficulty is rather that such a microstate, if present in an admissible ensemble and indefinite on $\beta$, counts neither positively nor negatively to the relative frequency of $\beta$. There is, however, an obvious way of dealing with this. We can still define an *interval* for the probability of $\beta$: a lower bound, on counting microstates in $\beta$, discarding all the indefinite cases (branches $P_\alpha|\psi\rangle$ that are not eigenstates of $P_\beta$); and an upper bound, on counting microstates in $\beta$ now including all the indefinite cases. So, if there are $n$ branches in all, of which $m$ are in $\beta$ whilst $r$ are indefinite on $\beta$, the probability of $\beta$ lies in $\left[\frac{m}{n}, \frac{m+r}{n}\right] \subset \mathbb{R}$. Call probabilities of this kind 'interval probabilities', as not rational numbers, but closed intervals of real numbers bounded by rational numbers. (We use intervals of reals rather than rational numbers purely for later convenience, in comparing them to Born-rule quantities.)

Probabilities of this general kind have been introduced before, but, it seems, mainly in connection with epistemic, agent-specific probabilities.[16] They are of the form $\mu_\mathbb{E}(\beta) = [p, q]$, $0 \leq p \leq q \leq 1$, where $p$ and $q$ are rational, as follows from their meaning as upper and lower bounds for relative frequency. Evidently

$$\mu_\mathbb{E}(\mathcal{M}) = [1,1]; \mu_\mathbb{E}(\emptyset) = [0,0],$$

follow from their definition. Addition and multiplication are defined in the obvious way:

$$[p, q] + [p', q'] = [p + p', q + q']$$
$$[p, q] \cdot [p', q'] = [p \cdot p', q \cdot q'].$$

Conditional probabilities for $\beta$ given $\alpha$, of the form $\mu_\mathbb{E}(\beta/\alpha)$, can also be defined for fixed ensemble $\mathbb{E}$, although more direct is to define new ensembles $\mathbb{E}'$ for the state $P_\alpha|\psi\rangle$, with the conditional given by $\mu_{\mathbb{E}'}(\beta)$. However, additivity of interval probabilities is not assured, in that $\beta, \beta'$ may be disjoint, yet

$$\mu_\mathbb{E}(\beta \cup \beta') \neq \mu_\mathbb{E}(\beta) + \mu_\mathbb{E}(\beta') \tag{6}$$

---

[15] In (Khawaja [2024]), it is suggested that a finest partitioning may be defined using a super-valuationist approach to the decoherence condition, with associated microstates those of non-zero amplitude. But even if this were to succeed in providing a cut-off in size of cells, it will hardly determine their shapes.

[16] An exception is (Walley and Fine [1982]), but they makes use of a propensity interpretation. The concept appears to go back to (Smith [1961]),( Good [1962]); see also (Huber [1973]). (My thanks to Jacob Barandes for alerting me to this literature.)

(see Fig.2b for an example). But there is a natural generalisation of additivity that does hold for any ensemble $\mathbb{E}$, namely:

$$\mu_{\mathbb{E}}(\beta \cup \beta') \subseteq \mu_{\mathbb{E}}(\beta) + \mu_{\mathbb{E}}(\beta'). \tag{7}$$

It is natural, because we are replacing probabilities as numbers by intervals; and it is a generalisation of additivity, as it reduces to additivity when $\mu_{\mathbb{E}}(\beta)$, $\mu_{\mathbb{E}}(\beta')$ are exact, with no Schrödinger-cat microstates for $\beta$ or $\beta'$. In fact, it reduces to additivity when there are no microstates in $\mathbb{E}$ which are superpositions of $\beta$ and $\beta'$: see Fig.2.

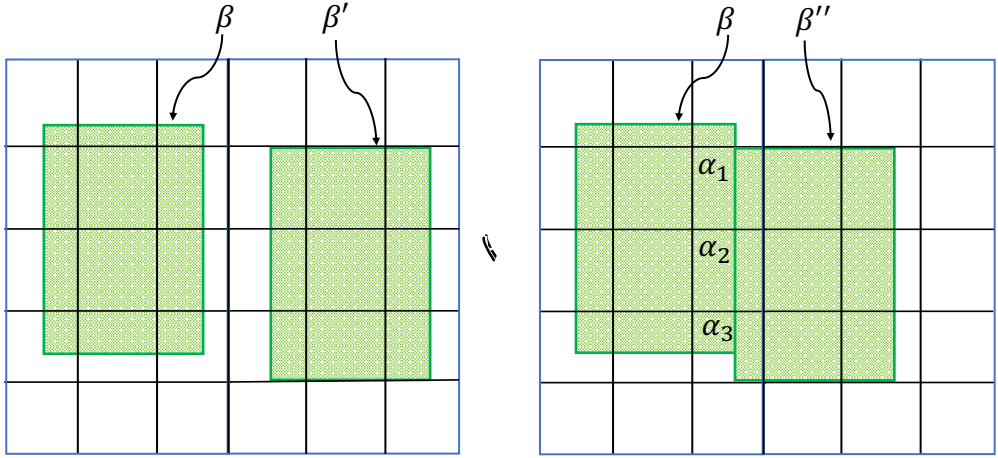

Figure 2a          Figure 2b

Microstates are represented by smaller rectangles ($n = 30$), with macrostates $\beta, \beta'$ stippled. $\mu_{\mathbb{E}}(\beta) = \frac{1}{n}[2,12]$, $\mu_{\mathbb{E}}(\beta') = \mu_{\mathbb{E}}(\beta'') = \frac{1}{n}[3,9]$. In Fig.2a there are several Schrödinger-cat microstates for $\beta, \beta'$, but no microstate is a superposition of both; $\mu_{\mathbb{E}}(\beta \cup \beta') = \frac{1}{n}[5,21] = \mu_{\mathbb{E}}(\beta) + \mu_{\mathbb{E}}(\beta')$. In Fig.2b three microstates are superpositions of $\beta$ and $\beta''$: $\alpha_1$ and $\alpha_2$ are now definite for $\beta \cup \beta''$, while $\alpha_3$ remains indefinite, but $\mu_{\mathbb{E}}(\beta \cup \beta'') = \frac{1}{n}[7,18] \subseteq \frac{1}{n}[5,21]$.

There is no requirement of additivity across different ensembles: Kolmogorov's axioms, and the generalised additivity property, hold for each ensemble considered separately. But as with exact partitionings, we demand consistency across ensembles. We read an interval probability for $\beta$ in $\mathbb{E}$ as defining bounds for the probability of $\beta$ simpliciter, without mention of an ensemble; we require that such bounds, for every admissible ensemble, *may all be jointly true* – requiring that they all have non-zero intersection. This is our second consistency condition: for any macrostate $\beta$, all interval probabilities, obtained on every admissible partitioning, must have non-zero intersection. If $\beta$ has an exact probability, that should be contained in that intersection too.

The two consistency conditions just adduced are similarly motivated and apply equally to the equiamplitude rule. Writing exact probabilities $p$ as $[p,p]$, we see that the former condition is a special case of the latter. And there is no reason to require consistency only for macrostates; we require the same for any region $\gamma$ of $\mathcal{M}$, macrostate or no.

In summary, and to keep the discussion as elementary as possible, stated in finitary terms:

**Consistency**: For any $\gamma \subseteq \mathcal{M}$, let $\mathcal{F}$ be any finite family of admissible ensembles. Then:

$$\bigcap_{\mathbb{E} \in \mathcal{F}} \mu_{\mathbb{E}}(\gamma) \neq \emptyset. \qquad (8)$$

Unlike generalised additivity, Eq.(7), which is satisfied automatically by both microstate counting rules, consistency in the sense of (8) is not. We consider the two rules by turn.

## 6. The equivolume rule

Let $\{\alpha_k\}$ partition $\mathcal{M}$ into three equivolume cells $\alpha_k$, with states $P_{\alpha_k}|\psi\rangle \neq 0$, $k = 1,2,3$. Consider the macrostate $\beta = \alpha_1$: it is exactly partitioned by $\{\alpha_k\}$, with probability 1/3. Consider the new partitioning $\{\alpha'_k\}$, with $\beta = \alpha'_1 = \alpha_1$, so that $\alpha_2 \cup \alpha_3 = \alpha'_2 \cup \alpha'_3$. Evidently at least one of $P_{\alpha'_2}|\psi\rangle$ and $P_{\alpha'_3}|\psi\rangle$ must be non-zero, since

$$P_{\alpha'_2}|\psi\rangle + P_{\alpha'_3}|\psi\rangle = P_{\alpha'_2 \cup \alpha'_3}|\psi\rangle = P_{\alpha_2 \cup \alpha_3}|\psi\rangle$$
$$= P_{\alpha_2}|\psi\rangle + P_{\alpha_3}|\psi\rangle \neq 0;$$

but there is no reason at all why both should be non-zero. For example, $|\psi\rangle$ might vanish outside of $\alpha'_1 \cup \alpha'_2$, even though non-zero for the complement of $\alpha_1 \cup \alpha_2$ (see Fig.3b). In that case, $\beta$ is exactly partitioned in a new ensemble containing just two states, $P_{\alpha'_1}|\psi\rangle$ and $P_{\alpha'_2}|\psi\rangle$, one of which is in $\beta$; so it has (exact) relative frequency 1/2, not 1/3. Contradiction.[17]

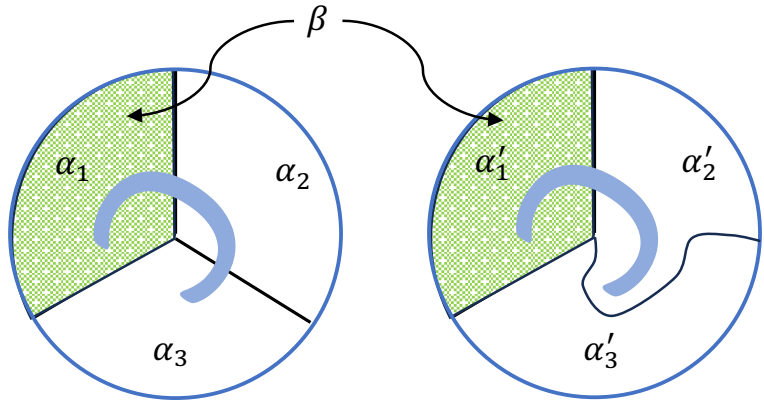

Figure 3a   Figure 3b

The support of the wave-function is the crescent shaded. in Fig.3a, macrostate $\beta$, stippled, coincides with cell $\alpha_1$; it is therefore exactly partitioned by $\{\alpha_1, \alpha_2, \alpha_3\}$, and since all cells have non-zero amplitude its probability is 1/3. In Fig.3b, $\beta$ is unchanged, and is again exactly partitioned by $\{\alpha'_1, \alpha'_2, \alpha'_3\}$, with $\alpha'_1 = \alpha_1 = \beta$; but now only $\alpha'_1$ and $\alpha'_2$ have non-zero amplitude, so $\beta$ has probability ½. Contradiction.

---

[17] The argument is in effect a synchronic version of the one mounted against Graham's branch-counting rule concerning diachronic consistency; see (Saunders [1998 p.388-9]), (Wallace [2012 p.120]).

The rule delivers inconsistent probabilities for the same total state and macrostate $\beta$, referred to different admissible ensembles. Can it be modified to escape this conclusion? Here is an alternative: apply the rule only to the region of $\mathcal{M}$ on which $|\psi\rangle$ (as a wave-function in $L^2(\mathcal{M}, d\lambda)$) is non-zero; that would seem to exclude the kind of counterexample just given.[18] The proposal is that the support of $|\psi\rangle$ on $\mathcal{M}$, the set of points $\mathcal{D}_{|\psi\rangle} = \{q \in \mathcal{M}; \langle q|\psi\rangle \neq 0\}$, is to be divided into equivolume cells, whereas before it was $\mathcal{M}$ that was so divided, with everything otherwise proceeding the same. Any cell wholly contained in the support of $|\psi\rangle$ on $\mathcal{M}$, of whatever size or shape, will then have non-zero amplitude, so branch-counting reduces to cell-counting in $\mathcal{D}_{|\psi\rangle}$. As such, it will be fully determined by the additive volume measure $\sigma$; plausibly, consistency is then ensured.

The suggestion is ingenious but fails on what looks like a technicality. The set of points $\mathcal{D}_{|\psi\rangle} \in \mathcal{M}$ will not in general be a connected sub-manifold of $\mathcal{M}$; excising points $\{q \in \mathcal{M}; \langle q|\psi\rangle = 0\}$ can leave as complicated a pattern of disconnected components imaginable. In general, no exact partitioning of even two components will be possible, for the same reason as before (see Fig.1). What then is the status of a cell $\alpha$ borderline, in this sense, on a component of $\mathcal{D}_{|\psi\rangle}$? Cells like this threaten to be ubiquitous. Since $\alpha \nsubseteq \mathcal{D}_{|\psi\rangle}$, by the amended rule, $P_\alpha|\psi\rangle$ does not count as a microstate in any ensemble, yet by assumption $\alpha \cap \mathcal{D}_{|\psi\rangle} \neq \emptyset$, so $P_\alpha|\psi\rangle \neq 0$, so it is not nothing, either. Before when we encountered borderline cells, they defined microstates indefinite on a macrostate; that did not impugn their status as microstates, as we took pains to explain. The new sort is more akin to indefiniteness as to *belonging to an ensemble*. For borderline $\alpha$, $P_\alpha|\psi\rangle$ is a superposition of a state belonging to an ensemble, and one not belonging to it -- but 'belonging to an ensemble' is not a macrostate.

Maybe sense can be made of the idea, maybe not; but if so, there remains the question of how cells borderline, in this new sense, are to contribute – or not contribute – to relative frequencies of macrostates. One strategy is to include them all, but that can immediately be dismissed, for it simply returns us to the previous rule – which, as we have just seen, is inconsistent. But to exclude all borderline cells also leads to inconsistency. In illustration, see Fig.4, where $\mathcal{D}_{|\psi\rangle}$ is the elliptical region shaded, and the rectangle in its interior is the macrostate $\beta$. Two equivolume partitionings are shown, yielding contradictory (exact) probabilities for $\beta$.

One could hope to use a mix of the two, and define a new kind of interval probability, with lower bound obtained by counting in all the borderline cells, the upper bound by excluding them, in both cases effecting the denominator. But with the rules for the numerators as before, the upper bound may well exceed unity. (Thus, in Fig 4a, suppose that $\beta$ is extended to $\beta'$ so that $\alpha_1$ is now borderline, with all else unchanged. By the original rule $\beta'$ will have interval probability [1/9,2/9]; by the new amendment, it is [1/9, 2/1].)

---

[18] My thanks to an anonymous referee for this suggestion.

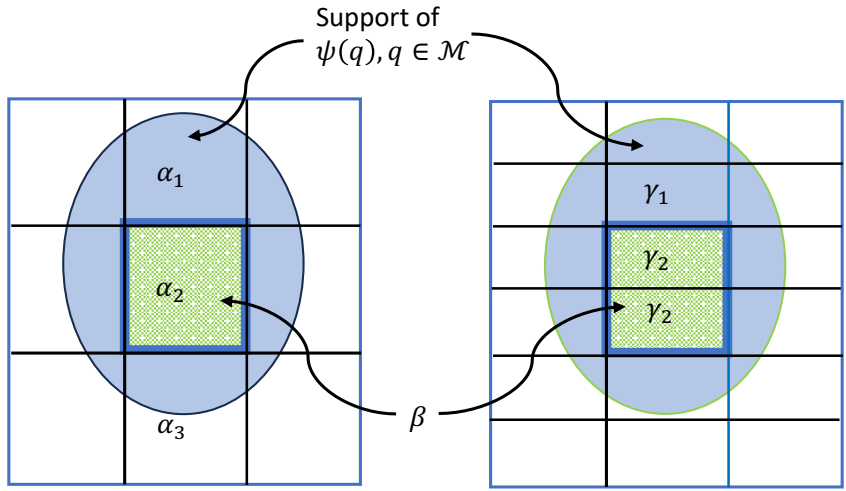

Figure 4a            Figure 4b

Two exact partitionings $\{\alpha_k\}, \{\gamma_k\}$ are shown for the same macrostate $\beta$, stippled; the support of $|\psi\rangle$ on $\mathcal{M}$ is shaded. In Fig 2a no cell disjoint from $\beta$ is contained in the support of $|\psi\rangle$, so the admissible ensemble consists of a single microstate $\alpha_2$ in $\beta$, and its probability is unity. In Fig.2b there is such a cell $\gamma_1$, with 2 microstates $\gamma_2, \gamma_3$ in $\beta$; the ensemble consists of 3 microstates so the probability of $\beta$ is 2/3, contradiction.

As a last resort, might we simply insist $\psi(q)$ be non-vanishing everywhere on $\mathcal{M}$, with the possible exception of sets of zero $\lambda$-measure? With that all these difficulties disappear. Take the original state $|\psi\rangle$, and add to it any analytic $L^2$-function on $\mathcal{M}$ with a sufficiently tiny amplitude: the resulting state can be as close as you like to the original, in the norm topology, yet be non-zero everywhere (except for sets of measure zero). But it is a pointless exercise. Combined with the non-zero amplitude condition, it ensures that the state is entirely irrelevant to the probabilities, for they will then be determined by $\beta$, $\mathcal{M}$, and $\lambda$ alone. In effect, it is to take the $\lambda$-measure as the probability measure, an admission of complete defeat.

## 7. The equiamplitude rule

We choose instead partitionings $\{\alpha_k\}$ in such a way that decohering microstates $P_{\alpha_k}|\psi\rangle$ all have equal amplitude, denote $\omega$. Since $\| |\psi\rangle \| = \sqrt{\langle\psi|\psi\rangle} < \infty$, it follows $\omega$ is restricted to the values $\omega_n := \| |\psi\rangle \|/\sqrt{n}$, where $n$ is an integer (the number of branches). We see again squares of amplitudes entering the picture, and with them square roots, traceable to the fact that Hilbert space is an inner-product space.

We should be sure, when $\mathcal{M}$ includes a continuous variable, that partitionings like this can be found. Consider again the case where $\mathcal{M}$ is a connected region in $\mathbb{R}^2$, equipped with Lebesgue measure. For any $n$, and any function $f(x,y) \in L^2(\mathcal{M}, dxdy)$, $\mathcal{M}$ can always be partitioned into connected regions $\alpha_k, k = 1, \dots, n$, where $\bigcup_k \alpha_k = \mathcal{M}$, such that

$$\int_{\alpha_1} |f|^2 dxdy = \int_{\alpha_2} |f|^2 dxdy = \cdots = \int_{\alpha_n} |f|^2 dxdy \,.$$

It follows that the functions $\chi_{\alpha_k}(x,y)f(x,y), k = 1,\ldots,n$, where $\chi_{\alpha_k}$ is the characteristic function of $\alpha_k$, are all orthogonal and have equal amplitude, where

$$f(x,y) = \sum_{k=1}^{n} \chi_{\alpha_k}(x,y)f(x,y)$$

is an expansion into $n$ equiamplitude states by construction. The argument generalises.

As with the equivolume rule, the larger the number of branches, the smaller the cells $\alpha_k$, so that eventually the decoherence condition will not be met. It follows that the number of equiamplitude branches cannot be taken as arbitrarily large. But as with the Boltzmann rule, there will b e uncountably many equiamplitude ensembles, for each $n$, as the above example shows.

The consistency condition Eq.(8) remains as before. Consider first the counter-example to the original Boltzmann rule for $n = 3$. Let $\mathcal{M}$ be partitioned into three cells $\{\alpha_k\}, k = 1,2,3$, as before, save now they have equal amplitude rather than equal $\lambda$-measure. It follows that $|P_{\alpha_k}|\psi\rangle| = \omega_3$. As before, we suppose that $\beta = \alpha_1$; therefore $\beta$ has (exact) probability 1/3. Let $\{\alpha'_k\}$ be a new equiamplitude partitioning into $n'$ branches of amplitude $\omega_{n'}$, with $\beta = \alpha'_1$, as before. Then $P_{\alpha_1}|\psi\rangle = P_{\alpha'_1}|\psi\rangle$, so their amplitudes are equal, so $\omega_{n'} = \omega_3$. Therefore $n' = 3$ and the relative frequency of $\beta$ in the new ensemble is 1/3. No counterexample for $n = 3$ is possible.

This proof also generalises, but there is another way to prove consistency of exact probabilities that is almost as simple, that extends to the inexact case. Suppose that for given $|\psi\rangle$ a macrostate $\beta$ is exactly partitioned by equiamplitude cells $\{\alpha_i\}$, $n$ in total, comprising ensemble $\mathbb{E}$. It follows as before that for some finite index set $K$:

$$P_\beta = \sum_{i \in K} P_{\alpha_i}. \qquad (9)$$

Let the cardinality of $K$ be $n_\beta$. Therefore, the relative frequency of $\beta$ is $n_\beta/n$. Since by assumption $\sum_{j=1}^{n} P_{\alpha_j}|\psi\rangle = |\psi\rangle$, from the definition of $\omega_n$ it follows:

$$\langle\psi|\psi\rangle = \langle\psi|\left(\sum_{i=1}^{n} P_{\alpha_i}\right)\left(\sum_{j=1}^{n} P_{\alpha_j}\right)|\psi\rangle = n\omega_n^2.$$

Further, from (9)

$$\langle\psi|P_\beta|\psi\rangle = \langle\psi|\left(\sum_{i=1}^{n} P_{\alpha_i}\right)\left(\sum_{k \in K} P_{\alpha_i}\right)\left(\sum_{j=1}^{n} P_{\alpha_j}\right)|\psi\rangle = n_\beta \omega_n^2.$$

Taking the ratio:

$$\frac{\langle\psi|P_\beta|\psi\rangle}{\langle\psi|\psi\rangle} = \frac{n_\beta}{n} = \mu_\mathbb{E}(\beta). \qquad (10)$$

The LHS is the Born rule quantity for $\beta$. The agreement is exact; no approximation is involved. The same exact agreement with the Born rule quantity follows for any other exact partitioning $\{\alpha_i'\}$ of $\beta$ with index set $K'$ and $n'$ branches (the argument is identical). Since the LHS of (10) is the same in each case, the RHS is too; therefore $n'_\beta/n' = n_\beta/n$ and $\mu_\mathbb{E}(\beta) = \mu_{\mathbb{E}'}(\beta)$.

In the case of a macrostate $\beta$ defined independent of any partitioning, a simple extension of this argument shows that the Born rule quantity for $\beta$ is contained within the limits of any (hence every) interval probability for $\beta$. That is, for any admissible ensemble $\mathbb{E}$ and any $\beta$:

$$\frac{\langle\psi|P_\beta|\psi\rangle}{\langle\psi|\psi\rangle} \in \mu_\mathbb{E}(\beta).$$

It follows that the intersection of interval probabilities for $\beta$ for any finite collection of admissible ensembles is non-empty, and the consistency condition is satisfied in full generality. The proof is in the appendix.

## 8. Arguments against finite frequentism

Stepping back, the Gibbs equiamplitude rule appears to be a fully reductive analysis, defined in terms of primitives that are widely used, in contexts that need have nothing to do with probability. The ensembles are actual (at least for Everettians) as well as finite, so unlike Gibbs' or Boltzmann's frequentism, and unlike the decision-theory approach to probability, it is clearly about the actual world (or rather worlds), independent of mind. It is inspired by actual scientific practise and theory – not just quantum mechanics, but classical statistical mechanics, in the very discovery of those theories. It is independent of metaphysics (unless realism about quantum states is metaphysical): it is independent, for example, of questions about personal identity, self-locating uncertainty, language use, or mind.

It appears, in short, to have all but one of the virtues of finite frequentism, as advertised by Hájek:

> Any aspiring frequentist with serious empiricist scruples should not give up on finite frequentism lightly. The move to hypothetical, frequentism say, comes at a considerable metaphysical price, one that an empiricist should be unwilling to pay. Finite frequentism is really the only version that upholds the anti-metaphysical, scientific inclination that might make frequentists of us in the first place. In any case, at first blush, it is an attractive theory. It is a reductive analysis, whose primitives are well understood; it apparently makes the epistemology of probability straightforward; unlike the classical and logical theories of probability, it appears to be about the world; and it seems to be inspired by actual scientific practice. (Hájek [1996 p.72]).

The exception is the epistemology of probability, which seems not at all straightforward (the topic of the final section). In compensation, there is one virtue that went unmentioned: relative frequencies, in a finite ensemble, classical or quantum, automatically satisfy Kolmogorov's axioms. If those are what chances are, the axioms are explained, including the generalised notion of additivity, in the case of interval probabilities. That is, our proposal is 'admissible' in Salmon's sense: the elementary axioms and their consequences follow from 'the meanings assigned to the primitive terms in the interpretation' (Salmon [1966, p.64]).

But having extolled the virtues of finite frequentism, Hájek went on to argue that it must fail – in 15 numbered arguments. The first and second were the most important and directly concerned the reference-class problem. I take it that problem, in the new proposal, has been comprehensively solved. The third and fourth (and to some extent the eighth) arguments involve the intuition that probabilities of local events should be counterfactually independent of distant events; so the ensembles relevant to the local event should likewise be local – yet typically no such local ensembles are observed. That problem too is solved: all the microstates in a superposition produced by a single local experiment exist at the same local place and time (and the dynamics explains why they cannot be observed). The fifth is similar: actual frequentism is ontologically over-burdened, because it implies, for an event to have an objective probability at all, that there must be multitudes of other events among which it is situated; the result is 'a surprisingly rich ontology'. Guilty as charged.

Argument 10 shows up an important difference in aim. It is the idea that a notion of objective probability may attach to statements about laws, or existence claims like 'there exist tachyons', and failing something like relative frequencies across possible worlds, it seems frequentism can give no leverage on these. Granted (and I'll not take the bait: the envisaged possible worlds are not among quantum mechanical branches). But we are seeking an interpretation of probability in terms of a physical theory, not proposing, as Hájek takes himself to be countering, a theory of objective probability to apply regardless of any particular physical theory. But oddly, we *are* proposing a universal theory of chance – but because quantum mechanics is a universal physical theory, not because the ideas apply to any physical theory.

Last but not least (except for the epistemological arguments) is Argument 14: finite frequentism can only ever deliver probabilities as rational numbers; what of continuously-valued probabilities, like the Born rule quantities?

At one level the answer is straightforward: we have already accounted for quantum probabilities, but as given by relative frequencies (and interval probabilities), not by the Born rule. True, the Born rule quantity for $\beta$ must always be contained in every interval probability $\beta$ (and equals any exact probability), but they are not themselves probabilities. The interval probabilities are perfectly intelligible in their own right; they do not need any underpinning as approximations to real-number valued probabilities.

A rather different answer is possible if we are prepared to allow that microstates need not be branches in Everett's sense; that they need not decohere. If would then be enough that they be orthogonal and written in the basis as defined by $\mathcal{M}$. In that case cells $\gamma \subseteq \mathcal{M}$ can be taken as arbitrarily small, and ensembles defined for any finite number $n$ of equiamplitude orthogonal microstates. The Born rule quantities can then be understood as the limits of interval probabilities – still not themselves relative frequencies, not themselves probabilities, but limits of such, as irrational numbers are limits of rational numbers.[19]

The price that is paid is that such microstates are no longer correlated with earlier ones, and that histories made up of microstates like these will strongly interfere with one another. They will no longer be dynamically autonomous, a criterion often used, as in Wallace [2012], for entities to count as worlds. Granted, then, they are not worlds. But they are still microstates, existing in a grand superposition, providing an explanatory statistical pattern (on which more in a moment).

---

[19] See (Stoica [2022]) for a related proposal.

One speculation quickly leads to another: why not extend the equiamplitude rule to *any* structure of the form $\langle \mathcal{M}, \lambda, |\psi\rangle \rangle$, where $\mathcal{M}$ is continuous? The probabilities defined in this way may not have any straightforward relation to experiment, but for realists, there need be nothing wrong with that. Indeed, if the basis can change from one ensemble to another there is no need for $\mathcal{M}$ to include a continuous variable. Equiamplitude decompositions always exist, for any state, in a finite-dimensional Hilbert space, if there is no constraint on the basis (before it was held fixed). (To take the example of a real 2-dimensional vector space, any vector can be written as the sum of two orthogonal vectors of equal length. This generalises.)

But if we are to take this route, and consider equiamplitude expansions of the state with respect to arbitrary bases, and in this way obtain relative frequencies for $\beta$ as defined in all these different ensembles, should we not insist on the same consistency condition as before? That is, in Eq.(8), the intersection of interval probabilities should be non-zero, now allowing for ensembles $\mathbb{E}$ of states in different bases.

And so it is. Consistency in the sense of (8) is ensured for different ensembles in different bases, for the same reason as for different ensembles in the same basis: because the Born rule quantity for $\beta$ is contained in any (hence every) admissible interval probability for $\beta$. For the proof, I refer to the appendix, noting the basis of the ensemble is arbitrary. Frequentism in this extended sense immediately implies non-contextualism, as probabilities are now defined for different bases (and the consistency condition is not assumed, it is proved).

## 9. Epistemology

The epistemology of probability in actual frequentism, in classical, one-world terms, is supposedly straightforward, whereas it is far from obvious for our ensembles of quantum microstates. But is it really so easy, classically?

It is simple for some kinds of empiricists -- those who, correctly insisting that only observed statistics provides an empirical basis for inferences about probabilities, wrongly identify the probabilities with those self-same statistics. Realists will insist that *all* extant statistics of the relevant sort, whether or not observed, determine the probabilities. So, in a classical world, chances are determined by ensembles of events spread out over all of time, or all time up to the present time, or all of space. It is to *those* relative frequencies that realists wish to infer, on the basis of local, observed data, for those are the real chances. But as soon as the observed ensemble differs from the inferred ensemble, the epistemology of probability seems far from straightforward.

In the quantum case, at first glance, it seems even worse. We face the apparently unnerving scenario that only a *single* microstate is ever observed, by any given observer (or local epistemic community), from the relevant ensemble. Ordinarily, with the ensemble spread out over space and time, we assume we (somehow) make multiple selections from that one ensemble at different places and times. But this comes at a certain price. The chance set-ups will not be precisely the same, on each trial; the ensemble is cobbled together out of several trials, only more or less physically similar, and of course, at most one outcome can be observed for each trial. It is the cobbled-together ensemble that is supposed to define the physical probability (in terms of statistics), one member for each trial; of which a sub-ensemble is actually observed, involving a place selection. Both target ensemble and observed ensemble involve these ambiguities.

In contrast, in the quantum case, the ensemble arises for each individual trial. The chance set-up for each member of the ensemble, prior to branching, is *identically* the same. No cobbling-together of approximately similar trials is needed to define the objective chances, for any single trial. But just because, for any given observer or community, only one element of the ensemble is observed, to obtain observed *statistics*, repeated trials at different places and times are needed, with outcomes recorded and made available at the end. In this way we are back to the same problem of cobbling together similar-enough chance set-ups, at our chosen places and times.

In brief: the epistemology of probability encounters much the same difficulties whether in one-world or in many-world frequentism, and is not at all straightforward on either account of probability; but it is somewhat simpler in the quantum case, where at least the objective chances are unambiguously defined.

This same distinction is important to answering Hájek's remaining arguments against finite frequentism, postponed until now. The sixth is that the actual relative frequencies produced by a series of trials may depart from the chances. An improbable sequence of outcomes is yet possible. A fair coin, tossed a dozen times and then destroyed, may yield a relative frequency of tails different from ½; tossed an odd number of times exact agreement is impossible. That is true in the quantum case too, but only concerning statistics built up of repeated trials, as for the observed relative frequencies. It is more than a possibility: necessarily, there will be 'anomalous branches', as they are called, in the Everett interpretation – branches in which the relative frequency of $E$ is different from the Born rule quantity for $E$. But they make up a vanishingly small proportion of branches, as the limit of the number of trials becomes large, hence with vanishingly small chance -- given that the trials are sufficiently similar.

A related argument is as follow (#7). Chance is supposed to explain relative frequencies. 'Why do we believe in chances? Because we observe that various relative frequencies of events are stable; and that is exactly what we should expect if there are underlying chances with similar values. But there is no explaining to be done if chance just is actual relative frequency: you can't explain something by reference to itself' (Hájek [1996 p.219]). The distinction between ensembles defined by repeated trials, and the ensembles produced on each trial, to the rescue again. The chance of $E$ for a given trial, is the relative frequency of $E$ in the microstates produced by that very trial. If the trials are sufficiently similar, the relative frequencies will be similar too. The observed statistics are stable (for the vast majority of observers – see the above); and what explains this stability is that the relative frequencies on each trial are indeed similar -- that the chances are the same on each trial. The explanans and the explanandum are different.

And so to the last of Hájek's arguments (#15): 'non-frequentist considerations enter our probabilistic judgments: symmetry, derivation from theory….'. Indeed. For our branch-counting rule, no sharp separation exists between the two, as symmetry and theory are involved from the beginning. The more helpful distinction is between probabilistic *judgment* (agent-specific, epistemic, or subjective probability, or credence), and the probabilities themselves (physical or objective probability, or chance). So suppose the relative frequency of $E$ in a (large) admissible ensemble is known: what credence should we form for $E$?

There is of course the question of what credence even *means*, in the face of branching, when all the relevant details are known in advance, but the answer to this question has been well-rehearsed in the decision-theory literature: regardless of one's views on personal identity, or self-locating uncertainty, or language use, agents confronted with different branching scenarios must still choose among actions, even if only whether to curl up in a ball and hide. These actions, even no action, involve the allocation of precious present resources -- and therein lies the meaning of credence. In the one-world case, for this calculus among competing actions to have any meaning, uncertainty about future scenarios with respect to existence is needed; for if there is none such, why consider scenarios that we know will not arise? In the many-world case, in contrast, it is precisely *because* we are certain of the existence of those competing future scenarios that we must take them into calculation – because we *know* they will all arise. They compete for our present resources, not for existence among themselves.

The question is then how, exactly, that calculation goes. It is clear what we want to end up with (and here I put interval probabilities to one side[20]): with Lewis' Principal Principle (roughly, that if agents know the objective probabilities for various scenarios, then their credence in those scenarios should be weighted accordingly). Of course, we could simply *use* the Principal Principle, as we do in conventional quantum mechanics, conforming credence in $E$ to its relative frequency, but we should aim higher: *why* ought an agent's credence in $E$ match the relative frequency of $E$ in an admissible ensemble?

I offer two answers. First, because it is prima facie reasonable. What better reason to give one eventuality more weight in our deliberations than another, when all else being equal, it is much more frequent than the other? That does not mean we should be careless of rare eventualities with very large negative utilities – again, just because those eventualities, though rare, are still there. (To those branches we funnel compensation, siphoned off from those not so afflicted, otherwise known as 'taking out an insurance policy'.) And all else surely *is* equal: there is an important symmetry operating in the equiamplitude rule, namely that an agent, prior to measurement, bears *exactly the same physical relations* -- in space, time *and* amplitude – to every element in every admissible ensemble produced by the measurement.[21]

My second answer is that it is reasonable that credence match relative frequency insofar as the axioms of the Deutsch-Wallace decision-theory approach are reasonable. For they show that agents who are reasonable in the latter sense will conform their credence to the Born rule quantities, so they will also conform to the relative frequencies, for the two agree.

However, those axioms are not entirely self-evident. Again, one may hope to do better. For at the core of the decision-theory derivations are just those symmetries of equiamplitude states, but now having to be realised at the level of *observably* distinct macrostates, that agents can bet on, involving loss or gain of rewards. Such are needed if credence is operationalised in terms of betting behaviour. This involves meticulous engineering of the experiment, depending on the precise initial state. Several of the axioms used by Deutsch and Wallace are directed to underwriting an agent's indifference to this kind of fine-tuning. But on the present account, following a measurement (any measurement), equiamplitude microstates, albeit differing in unobservable ways, are already there for the taking, along with their symmetries, with no contrived experiments needed to reveal them. It may be they can be exploited more directly.[22]

---

[20]There is a large literature on how credence works using interval probabilities, then often called 'imprecise probabilities' (Bradley [2019]), but in practise, for ensembles produced in quantum measurements, the interval widths can be taken as extremely small (much smaller than experimental uncertainties in the values of the Born rule quantities).

[21] Differences in phase can be reduced to phase relations among the branches, not between the agent prior to branching and the individual branches (because of the irrelevance of the overall phase).

# Acknowledgments

My thanks to Harvey Brown, and to two anonymous referees, for extremely helpful suggestions and constructive criticism.

# Appendix

**Theorem**. For any projection operator $P_\beta$ on $\mathcal{H}$ and any ensemble $\mathbb{E}$ of orthogonal states $|\varphi_j\rangle \in \mathcal{H}$ of equal amplitude, where $\sum_j |\varphi_j\rangle = |\psi\rangle$:

$$\frac{\langle\psi|P_\beta|\psi\rangle}{\langle\psi|\psi\rangle} \in \mu_E(\beta).$$

**Proof**. Let $\mathbb{E}$ have $n$ elements, each with amplitude $\omega_n = \|\psi\rangle\|/\sqrt{n}$. Let $m \leq n$ satisfy $P_\beta|\varphi_j\rangle = |\varphi_j\rangle$, and denote such states $|\phi_j\rangle$; let $n - m - r \geq 0$ satisfy $P_\beta|\varphi_j\rangle = 0$, denote $|\eta_j\rangle$, and let $r$ satisfy neither, denote $|\xi_j\rangle$. The interval probability for $\beta$ in ensemble $\mathbb{E}$ is therefore:

$$\mu_E(\beta) = \frac{1}{n}[m, m+r].$$

The total state expanded in the new notation is:

$$|\psi\rangle = \sum_{j=1}^{n}|\varphi_j\rangle = \sum_{j=1}^{m}|\phi_j\rangle + \sum_{j=1}^{r}|\xi_j\rangle + \sum_{j=1}^{n-m-r}|\eta_j\rangle$$

where all states have equal amplitude. Substituting in $\langle\psi|P_\beta|\psi\rangle$, we obtain six expressions, only two of which are non-vanishing. Three are sums of cross terms between states of different types, of the form

$$\langle\xi_j|P_\beta|\phi_k\rangle; \ \langle\phi_j|P_\beta|\eta_k\rangle; \ \langle\eta_j|P_\beta|\xi_k\rangle.$$

They all vanish by inspection. Likewise sums of terms $\langle\eta_j|P_\beta|\eta_k\rangle$ vanish. For the expression containing only $|\phi_j\rangle$'s we obtain

---

[22] I make some brief suggestions on how this may go in (Saunders [2021]). (Short [2023]), viewed as a derivation of an agent's credence function rather than physical probability, is a further step in this direction (with the 'invariance' condition following from frequentism).

$$\langle \varphi_1 + \cdots + \varphi_m | P_\beta | \varphi_1 + \cdots + \varphi_m \rangle = m\omega_n^2.$$

The remaining term involving only states indefinite on $\beta$ is

$$\langle \xi_1 + \cdots + \xi_r | P_\beta | \xi_1 + \cdots + \xi_r \rangle.$$

It is positive definite and bounded above by $r\omega_n^2$. Therefore for $r > 0$, $m\omega_n^2 < \langle \psi | P_\beta | \psi \rangle < r\omega_n^2$; when $r = 0$, $m\omega_n^2 = \langle \psi | P_\beta | \psi \rangle$, and the result follows. Since $n$ and $\mathbb{E}$ were arbitrary, the conclusion is quite general. ∎

Faculty of Philosophy,
University of Oxford
Radcliffe Humanities,
Oxford OX2 6GG,

Merton College,
 Oxford OX1 4JD